# Performance Study of ETX based Wireless Routing Metrics

Nadeem Javaid[1], Akmal Javaid[3], Imran Ali Khan[4], Karim Djouani [1,2]

[1]LISSI, University of Paris Est, France, Email: nadeem.javaid@etu.univ-paris12.fr
[2]F'SATIE, TUT, South Africa, Emails: djouanik@tut.ac.za, djouani@univ-paris12.fr
[3]CIIT, Wah Cantt, Pakistan, Email: akmaljavaid@ciit-wah.edu.pk
[4]CNIC, Chinese Academy of Sciences, Beijing, China, Email: imran@cnic.cn

*Abstract*—Being most popular and IETF standard metric, *minimum hop count* is appropriately used by *Ad hoc Networks*, as new paths must rapidly be found in the situations where quality paths could not be found in due time due to *high node mobility*. There always has been a tradeoff between throughput and energy consumption, but *stationary topology* of *WMNs* and *high node density* of *WSN's* benefit the algorithms to consider quality-aware routing to choose the best routes. In this paper, we analytically review ongoing research on wireless routing metrics which are based on ETX (Expected Transmission Count) as it performs better than minimum hop count under *link availability*. Performances over ETX, target platforms and design requirements of these ETX based metrics are high-lighted. Consequences of the criteria being adopted (in addition to expected link layer transmissions & retransmissions) in the form of incremental: (1) performance overheads and computational complexity causing inefficient use of network resources and instability of the routing algorithm, (2) throughput gains achieved with better utilization of wireless medium resources have been elaborated.

## I. INTRODUCTION

On a wireless link, the number of link layer transmissions of a packet is an appealing cost metric because minimizing the total number of transmissions (and retransmissions) maximizes the throughput of an individual link then overall network. *ETX* proposed in [6], measures MAC transmissions and retransmissions to recover from frame losses since the link level retransmissions depend only on the link level packet errors caused by channel issues. *ETX* of a wireless link is the estimated average number of transmissions of data frames and ACK frames necessary for the successful transmission of a packet [5]. A node derives *ETX* by estimating the frame loss ratio at the link $l$ to each of its neighbors in the forward direction as $p_{lf}$, and in the reverse direction as $p_{lr}$ transmitting broadcast probe packets (which are not retransmitted) at the link layer once every second as:

$$ETX_l = \frac{1}{(1-p_{lf})(1-p_{lr})} \quad (1)$$

Alternatively, *ETX* of the link is the inverse of the probability of "successful packet delivery" or "link reliability":

$$ETX_l = \frac{1}{p_{lsf} * p_{lsr}} = \frac{1}{reliability(l)} \quad (2)$$

If we increase the frequency of *ETX* measurements and change the optimum paths accordingly more frequently, it involves significant amount of overhead in the network. It has been shown that the link with a lower *ETX* metric may in fact lead to a higher observed loss rate at the transport layer. Because good link-layer protocols do not retransmit lost packets forever and give up after a threshold number of attempts. The losses occurring in the form of bursts cause to pick the link in the middle of a burst-error situation, which is bad even with a lower *ETX*.

Consider for example, the Fig.1, which illustrates the packet delivery ratios taken from four distinct links in the Roofnet wireless mesh network [2]. Each of these four links has an *ETX* around 2 during the testing period. Therefore, if *ETX* is taken as the metric for quality, these four links are identical. On the other hand, the sample variances of the delivery ratios are quite different for these links, i.e., these wireless links have similar long term average behaviors, even though their short-term behaviors are quite different [3].

Pithily, *ETX* does improve the throughput of a wireless network (with less mobility) when compared to hop count metric but it does not track the variations on the channel at short time scales due to potential route instability [1]. Table 1 lists the performances over *ETX*, design goal and experimented platforms of the *ETX* based metrics.

## II. STUDY OF ETX BASED METRICS

### 1. Modified ETX (mETX) and Effective Number of Transmissions (ENT)

In almost all kinds of wireless networks, due to the fast link-quality variation, the metrics based on a time-window interval, such as *ETX, ETT, WCETT, MIC, MCR, iAWARE*, etc., may not follow the link quality variations and/or may produce prohibitive control overhead. To cope with the situation, *mETX* and *ENT* were proposed in [3], which are aware of the probe size, therefore, the inclusion of the data rate is trivial for them. Along with the link-quality average values, these metrics consider the standard deviation to project physical-layer variations.

## 1.1. mETX

Presence of *channel variability* in *ETX* became the reason to design *mETX*. The difference between *mETX* and *ETX* is: rather than considering probe losses, *mETX* works at the bit level. The *mETX* metric computes the bit error probability using the position of the corrupted bit in the probe and the dependence of these bit errors throughout successive transmissions. This is possible because probes are composed by a previously known sequence of bits. The variability of the link is modeled using the statistics of this stochastic process. Then, the mean number of transmissions is analytically calculated and the results show that it can be closely approximated with the statistics of the bit error probability, summed over a packet duration. For *mETX*, the critical time scale for the link variability is the transmission time of a single packet including all its retransmissions. *mETX* is defined in eq.(3) with μ being the estimated average packet loss ratio of a link and $\sigma^2$ the variance of this value. Like *ETX*, *mETX* is additive over concatenated links.

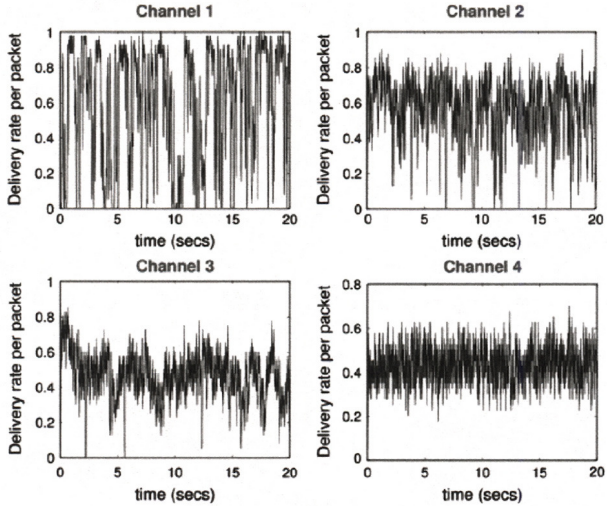

Fig.1. Roofnet, Packet delivery rate for four distinct links

$$mETX = \exp\left(\mu + \frac{\sigma^2}{2}\right) \quad (3)$$

μ means the impact of slowly varying and static components, like shadowing, slow fading in the channel and $\sigma^2$ shows the impact of relatively rapid channel variations, as fading, interference, etc. which the term $\mu_\Sigma$ (and hence the *ETX*) cannot track. μ and $\sigma^2$ are estimated by the bit level information, as counting only the packet losses is not sufficient; thus, parameters μ and $\sigma^2$ are estimated by considering the number of erred bits in each probe packet.

Complexity of "channel estimation" is the main disadvantage of the *mETX*, as: (1) probe packets are to be processed at the bit level raising energy consumption issue in wireless sensor networks (which may not be an issue for wireless mesh networks due to their abundance of processing power), (2) $\sigma^2$ increases with increased estimation error. A link's high *mETX* is due to high channel variability and estimation error which results a better link to be malformed. *mETX* can be adapted like *ETX* easily for those wireless links which provide bit rate adaptation by normalizing the metric according to the transmission rate.

## 1.2. Effective Number of Transmissions (ENT)

The upper-layer protocols, as TCP, SPX, etc. have a limit to tolerate (re)transmissions. This issue caused *ENT* to be proposed. *ENT* therefore, broadcasts probes, limits routes to an acceptable number of (re)transmissions according to the requirements of upper-layer. It measures the number of successive retransmissions per link considering the variance to find a path that achieves high network capacity while ensuring that the end-to-end packet loss rate visible to higher layers (such as TCP) does not exceed a specified value but this may not be sufficient, as it may involve links with high loss rates. *ETX* and *mETX* metrics usually select the links which do not obey the transmission threshold required by the algorithms working at higher layers.

Let *M* be the threshold number of retransmissions (specified by higher layers), $P_{al}$ (actual probability of a packet loss), using a large deviations approach can be defined as:

$$P_{al} \approx exp\left[-(logM - \mu)^2/2\sigma^2\right] \quad (4a)$$

*G* be the temporal diversity gain for a wireless link:

$$G = -\ log P_{dl}/log\ M \quad (4b)$$

which specifies the desired loss probability $P_{dl}$. Now, *ENT* can be defined as:

$$ENT = \exp\left(\mu + 2G\sigma^2\right) \quad (4c)$$

If higher layer does not specify any loss probability constraint, i.e., $G = 0$, then for the given $P_{dl}$, $\mu + 2G\sigma^2 \leq \log M$, we left with $\mu \leq \log M$. If $G > 0$ then there must be efficient resources for the network to put an amount $2G\sigma^2$ (which is directly related to the variability of the channel, $\sigma^2$ and strictness of the loss requirement, *G*). This interpretation of *ENT* is analogous to the notion of effective bandwidth, which was developed to model variable traffic sources in queuing networks. Indeed, *ENT* can be interpreted as the effective bandwidth of the discrete stochastic process, the number of transmissions.

Differences between the two include: (1) an extra degree of freedom due to the factor 2G in *ENT*. Indeed (*mETX* is the *ENT* evaluated at $\frac{1}{4}$), (2) *ENT* is not additive as *ETX* or *mETX*. Similarities between the two include: (1) a by-product of *ENT* reduces the packet loss ratio observed by higher-layer protocols, after any link-layer retransmissions are done, (2) they have same channel estimation procedure. Main feature of *ENT* is that it can be calibrated. It is useful to have a degree of freedom for the necessary adjustments derivations in [3] are based on certain assumptions, which can be partly violated in different platforms and environments. Both of drawback of *mETX* metric stated in section 1.1 are valid for the *ENT* as well.

## 2. Expected Transmission Time (ETT)

*ETT* is the time a data packet needs to be successfully transmitted to each neighbor. To overcome *ETX's* shortcomings: (1) it broadcasts at the network basic rate, (2) its probes are smaller than data packets, *ETT* [4] adjusts *ETX* to different PHY rates and data-packet sizes. Two approaches to compute the bandwidth of link $l$, $B_l$ are:

Eq.(1) from [4] can be re-written as:

$$ETT_l = ETX_l * t \qquad (5)$$

$$ETT_l = ETX_l * \frac{S_F}{B_l} \qquad (6)$$

$$ETT_l = ETX_l * \frac{S_F}{S_L/(T_S - T_L)} \qquad (7)$$

Where $S_F$ is the data packet of fixed-size, $B_l$ is bandwidth of link $l$, $S_L$ is data packet of largest-size, $T_S - T_L$ is an interval between the arrivals of two packets. This technique unicasts two packets in sequence, a small one followed by a large one, to estimate the link bandwidth to each neighbor measuring the inter-arrival time period $T_S - T_L$ between the two packets and reporting it back to the sender. The computed bandwidth is the size of the large packet of the sequence divided by the minimum delay received for that link.

Eq.(2) from [6], i.e., loss probability is estimated by considering that IEEE 802.11 uses data and ACK frames. Loss rate of data is estimated by broadcasting a number of packets of the same size as data frames, one packet for each data rate defined in IEEE 802.11. Loss rate of ACK frames is estimated by broadcasting small packets, of the same size as ACK frames and sent at the basic rate, which is used for ACKs. *ETT* may choose a path that only uses one channel, even though a path with more diversified channels has less intra-flow interference and hence higher throughput. Similarly to *ETX*, the chosen route is the one with the lowest sum of *ETT* values.

## 3. Weighted Cumulative ETT (WCETT)

Basically *WCETT* is based on *ETT* and aware of the loss rate (due to *ETX*) and the bandwidth of the link. Proposed by [8], *WCETT* can be used for multi-radio, multi-hop WMN. It proposes *ETT* which improves on *ETX* by making use of the data rate in each link. The *ETT* of a link is defined in (6). *ETT* explains the expected MAC transmission time of a packet of a size *S* over certain link *l*. Given the presence of multiple channels and intra-flow interfere, *WCETT* is defined as:

$$WCETT = (1 - \beta) + \sum_{l=1}^{n} ETT_l + \beta * \max_{1 \leq j \leq k} X_j \qquad (8)$$

*WCETT* is the sum of *ETT*'s of all the links in path *p* operating on $X_j$ channel *j*, in a system with total of *k* orthogonal channels. $\beta$ is a tunable parameter subject to $0 \leq \beta \leq 1$. *WCETT* consists of two components: the first component finds the path with the least sum of *ETT*'s; the second accounts for the bottleneck channel dominating the throughput of the total path.

Its advantages include: (1) over performing *ETT*, it explicitly accounts for the intra-flow interference, providing support for multi-radio or multi-channel wireless networks, (2) its two weighted components of it substitute the simple summation of *ETT* and attempt to strike a balance between throughput and delay. It does not capture inter-flow interference compared with Interference Aware Routing Metric (*iAWARE*). It modifies *ETT* considering intra-flow interference. This metric is a sum of end-to-end delay and channel diversity. Like Minimum Loss (*ML*) and unlike *ETX* & *ETT*, *WCETT* is an end-to-end metric because it must consider all channels used along the route to avoid intra-flow interference.

## 4. Metric of Interference and Channel-switching (MIC)

*WCETT* does avoid intra-flow interference but it does not (1) guarantee shortest paths (2) avoid inter-flow interference; which may lead *WCETT* to select congested routes. *MIC* [9], [10] tackles these issues by providing the features: (1) each node estimates inter-flow interference by counting the number of interfering nodes in the neighborhood. (2) *MIC* virtual nodes guarantee minimum-cost routes computation. (3) *MIC* calculates itself by *ETT* metric. *MIC* for a path *p* is defined as follows:

$$MIC(p) = \frac{1}{N \times min(ETT)} \sum_{l \in p} IRU + \sum_{i \in p} CSC_i \qquad (9)$$

Where *N* is the total number of nodes in the network and $min(ETT)$ is the smallest *ETT* in the network. The two components of *MIC*, *IRU* (Interference-aware Resource Usage) is $IRU_l = ETT_l \times N_l$ and *CSC* (Channel Switching Cost) is defined as:

$CSC_i = w_1$; if $CH(prev(i)) \neq CH(i)$ and

$CSC_i = w_2$; if $CH(prev(i)) = CH(i)$. Where $0 \leq w_1 < w_2$, and *N* is the set of neighbors that interfere with the transmissions on link *i*. $CH(i)$ represents the channel assigned for node *i*'s transmission and $prev(i)$ represents the previous hop of node *i* along the path *p*. *MIC* takes the inter-flow interference into account. Its disadvantages include: (1) the component, *CSC* captures intra-flow interference only in two consecutive links. (2) *MIC* considers interference of a link caused by each interfering node in the neighborhood, counts the amount of interferers on a link only by the position of the interfering nodes no matter whether they are involved in any transmission simultaneously with that link. *MIC*, therefore, utilizes the measurement of signal power to capture inter-flow and intra-flow interference.

## 5. Interference AWARE (iAWARE)

*iAWARE* considers not only both inter-flow and intra-flow interference and characterized by the physical interference model but also takes link-quality variation into account. This metric uses Signal to Noise Ratio (SNR) and Signal to Interference and Noise Ratio (SINR) to continuously reproduce neighboring interference variations onto routing metrics. The *iAWARE* metric estimates the average time the medium is busy because of transmissions from each interfering neighbor. Higher the interference, higher the *iAWARE* value. Thus, unlike *mETX* and *ENT*, *iAWARE* considers intra-flow and inter-flow interference, medium instability, and data-transmission time. In this model [11], a communication between nodes *u* and *v* on the link *(u→v)* is successful if the *SINR* at the receiver *v* is above a certain threshold. Let $P_u(v)$ denotes the signal strength of a packet from node *u* to node *v*. *iAWARE*'s first component, finds paths with least path cost and other finds paths with least intra-flow interference (exploiting channel diversity). Moreover, the introduction of *SINR* is a great breakthrough for inter-flow interference-aware routing compared with other *ETX*-based metric, like *MIC*.

Define of the link metric *iAWARE* of a link *j* as follows:

$$iAWARE_j = \frac{ETT_j}{IR_j} \qquad (10a)$$

When $IR_j$ for the link *j* is 1 (no interference), $iAWARE_j$ is simply $ETT_j$ which captures the link loss ratio and packet transmission rate of the link $j$. $ETT_j$ is weighted with $IR_j$ to capture the interference experienced by the link from its neighbors. A link with low $ETT$ and high $IR$ will have a low $iAWARE$ value. Lower the $iAWARE$ of a link better is the link. We define interference ratio $IR_i(u)$ for a node $u$ in a link $i = (u, v)$, where $IR_i(u)$ $(0 < IR_i(u) \leq 1)$ can be defined as:

$$IR_i(u) = \frac{SINR_i(u)}{SNR_i(u)} \quad (10b)$$

$$SNR_i(u) = \frac{P_u(v)}{N} \quad (10c)$$

$$SINR_i(u) = \frac{P_u(v)}{[N + \sum_{w \in \eta(u)-v} \tau(w)P_u(w)]} \quad (10d)$$

Here $\eta(u)$ denotes the set of nodes from which node $u$ can hear (or sense) a packet and $\tau(w)$ is the normalized rate at which node $w$ generates traffic averaged over a period of time. $\tau(w)$ is 1 when node $w$ sends out packets at the full data rate supported. We use $\tau(w)$ to weight the signal strength from an interfering node $w$ as $\tau(w)$ gives the fraction of time node $w$ occupies the channel.

## 6. Distribution Based Expected Transmission Count (DBETX)

Through a complete physical channel view and using cross-layer optimizations, Distribution Based Expected Transmission Count (DBETX) is proposed in [7] to improve network performance for varying channels and in the presence of fading.

DBETX's performance over ETX increases with the network density because connectivity increases and more routing options become available. Results show a reduction of up to 26% in the Average Number of Transmissions (ANT) per link and an increase of up to 32% in the end-to-end availability.

Using link measurements, DBETX makes the nodes able to: (i) estimate the probability density function (pdf) of the experimented SNIR, (ii) calculate the expected Bit Error Rate (BER) and, as a consequence, the expected packet error rate (PER), (iii) estimate average number of required transmissions in a given link based on the SNIR, (iv) derive the number of required transmissions taking into account the maximum number of MAC-layer retransmissions, (v) penalize lossy links in order to find routes with lower end-to-end loss rates, (vi) reflect the variations of the wireless channel, (vii) to favor links with a lower loss probability (oppositely from [3]). DBETX metric for a link is defined as:

$$DBETX(l) = E[ANT(l)] * \frac{1}{1 - P_{out_{MAC}}} \quad (11a)$$

$P_{out_{MAC}}$ is the probability when $P_{Suc}(x) < P_{limit}$. Where $ANT(l)$ is given by:

$$ANT(l) = \frac{1}{P_{Suc}(x)}; \quad P_{Suc}(x) > P_{limit} \quad (11b)$$

$$ANT(l) = \frac{1}{P_{limit}}; \quad P_{Suc}(x) \leq P_{limit} \quad (11c)$$

MAC layer outage (a condition when current Success Probability ($P_{Suc}$) of a link results in an expected number of retransmissions higher than MaxRetry) occurs when the success probability of a link is smaller than the Limit Success Probability ($P_{limit}$), which is $P_{limit} = 1/MaxRetry$. In this situation, there is a high probability that the transmitted packet will be discarded due to an excessive number of retransmissions. ANT function is the expected number of retransmissions on a link considering the value of MaxRetry, which is the maximum number of retransmissions allowed by the MAC-layer (For IEEE 802.11, it is 7 in the presence of RTS/CTS handshake). DBETX's calculation requires the information of actual behavior of wireless link instead of the average behavior. Due to the difference in the working time scales of the different layers, it is impossible to have a complete view of physical medium based on network level, as the events of interest occur at the physical level at milli or microseconds, network level interactions are reduced in order to reduce overhead at a time scale of seconds.

## 7. Exclusive Expected Transmission Time (EETT)

In large-scale multi-radio mesh networks (LSMRMNs), most of traffic has much longer paths than in small scale WMNs [19]. When channels are distributed on a long path, EETT selects multi-channel routes with the least interference to maximize the end-to-end throughput. None of existing routing metrics is capable to evaluate two multi-channel paths accurately when the paths are long. So, EETT well considers channel distribution on long paths which however are very critical in LSMRMNs. In order to meet the above mentioned requirement, EETT is used to give a better evaluation of a multi-channel path. For a $N$-hop path with $K$ channels, on a link $l$, its Interference Set (IS) is the set of links that interfere with it (a link's IS also includes the link itself). Then this link $l$'s EETT is defined as:

$$EETT_l = \sum_{l \in IS(l)} ETT_l \quad (12)$$

Physical interpretation of EETT states that EETT of a link $l$ shows the channel used by link $l$. Link $l$ may have to wait a longer period for transmission on a channel, if there are more neighboring links on that channel with link $l$ resulting in a path with a larger EETT with more severe interference and needs more time to finish the transmission over all links within the path. EETT reflects the optimality of the channel distribution on a path, as this results in less intra-flow interference. EETT can also embody the inter-flow interference, if IS(l) includes those links which do not belong to the same path with link $l$. MIC considers the impact of link $l$ on other links, while EETT considers the impact of other links on link $l$ hence EETT is supposed to have better performance since it more accurately reflects the impact of the inter-flow interference.

## 8. Expected Data Rate (EDR)

To overcome ETX's key limitation of not taking into account the multi-rate links, ETT was proposed to account for multi-rate links. Transmission Contention Degree, (TCD) in [12] was defined to overcome the limitation of ETX and ETT for making conservative estimates for paths longer than 3-4 hops (as all the cochannel links on a path contend with each other) by incorporating time-sharing effects of MAC. TCD is the average fraction of the time for which the outgoing queue of the transmitter of link $l$ is non-empty. EDR is defined as:

$$EDR_l = \frac{b_l}{ETX_l * \sum_{i=1}^{n} TCD_l(i)} \quad (13)$$

Where $b_l$ is the nominal bit rate of the link $l$. $\sum_{i=1}^{n} TCD_l(i)$ is used to account for throughput reduction due to *equal* time-sharing with the contending links provided that all the links have the same nominal bit rate. If links have different nominal bit rates, they receive the same average throughput, but different time-share of the channel failing to capture the bandwidth-sharing mechanism of 802.11 DCF.

### 9. Expected Throughput (ETP)

Being proposed in [20], *ETP*: (1) predicts better routes than ETX and ETT in mesh networks with long paths, they do not make spatial measurements, (2) also measures expected throughput of a link, (3) can easily be implemented in the *IBSS* mode with minor additions to the beacon message contents, (4) predicts better routes in mesh networks with heterogeneous link rates because *ETP* captures the bandwidth sharing mechanism of 802.11 DCF more accurately than *EDR*, *ETT*, and *ETX*, as they do not take into account the throughput reduction of fast links due to contention from slow links. (5) *ETP* is suitable for multi-rate, multi-radio mesh networks.

To state *ETP*, let link *l* belongs to path *P* in the contention domain $S_l$. $S_l \cap P$ is the set of links on path *P* that contend with link *l*. $r_l$ be the nominal bit rate of link *l*. All links have equal number of opportunities for transmission when saturated, as per 802.11 DCF. The expected bandwidth received by each link *l* is:

$$b_l = \frac{1}{\sum_{j \in S_l \cap P} \frac{1}{r_j}} \quad (14a)$$

But the packet losses lower the actual throughput of the link. $p_l^{(f)}$ and $p_l^{(r)}$ are supposed to be the packet success probabilities of link *l* in the forward and reverse directions respectively, then the *ETP* of link *k* is given by:

$$ETP_l = \frac{p_l^{(f)} * p_l^{(r)}}{b_l} \quad (14b)$$

In the form of *ETX*, we have:

$$ETP_l = \frac{1}{ETX_l * b_l} \quad (14c)$$

i. e., it is computing the expected throughput of a link directly. *f(P)*, is the throughput of the bottleneck link of the path,

$$f(P) = \min_{k \in P} ETP(k) \quad (14d)$$

Unlike *ETX, ETT* and *EDR, ETP* has a more accurate model for the impact of contention in 802.11 MAC.

### 10. Multi-channel Routing Protocol (MCR)

*WCETT* lacks switching cost so [13] added it in (8) and suggested *MCR* as:

$$MCR = (1 - \beta) \times \sum_{l=1}^{n} (ETT_l + SC(c_l)) + \beta * \max_{1 \leq j \leq c} X_j \quad (15)$$

The additional component, Switching Cost, $SC(c_l)$ is defined as follows:

$$SC(c_j) = \sum_{\forall i \neq j} InterfaceUsage(i) \times SwitchingDelay$$

This value does not figure in the time interval that this interface is tuned to channel *j*, but is idle. *SwitchingDelay* is latency for switching an interface and can be measured offline. When a packet arrives on channel *j*, $\sum_{\forall i \neq j}^{n} InterfaceUsage(i)$ measures the probability that the switchable interface will be on a different channel ($i \neq j$).

Like *WCETT, MCR* fails to figure the inter-flow interference besides the assumption that all available channels are orthogonal but channel-switching cost makes *MCR* to be incorporated with the routing protocol like DSR, AODV for multi-channel and channel-switchable wireless network.

### 11. Medium Time Metric (MTM)

*MTM* [14] minimizes the time of consumption of physical medium. Due to the shared nature of wireless networks, not only individual links may interfere (intra-flow interference) but transmissions compete for the medium with each other in the same geographical domain. The longer the physical distance of a hop results in the higher energy consumption and the more other hops are affected. The *MTM* of a packet *p* on a path *P* is defined as follows:

$$MTM(P, p) = \sum_{\forall l \in P} \tau(l, p) \quad (16)$$

Where $\tau(l, p)$ is the time required to transfer packet *p* over link *l*. $\tau(l, p)$ is defined as:

$$\tau(l, p) = \frac{overhead(l) + \frac{size(p)}{rate(l)}}{reliability(l)} \quad (17a)$$

Utilizing (2) and (6), (17) in the form of *ETX* is:

$$\tau(l, p) = ETX(overhead(l) + t) \quad (17b)$$

Link overhead can be computed from standards and specifications as well as from the type and configuration of the used wireless device. The packet size should be easily available through the routing protocol. Link transfer rate and reliability usually are known to the MAC layer. However, this information often is not accessible to higher network layers because the techniques used for auto-rate selection on the MAC layer are considered proprietary. It is possible to estimate the values for transfer rate and link reliability by probing. Though, this information produces unnecessary overhead and less accurate results than inter-layer communication would.

Therefore, Awerbuch *et al.* [8] would favour that radio card manufacturers provide a standard interface in order to enable access to this information by higher network layers. Although we agree with them principally, one should not expect that all problems of measuring transfer rate or link reliability be solved at once thereby. Awerbuch *et al.* [8] measured an end-to-end throughput which was equal to minimum hop count and *ETX* in short distances. When the distances were larger, minimum hop count and *ETX* found routes with a few hops. *MTM* selected multi-hop paths with more hops but higher capacity. For this reason, the resulting end-to-end throughput was up to 20 times higher with *MTM* than with the other metrics.

### 12. Estimated Transmission Time (EstdTT)

Neglecting the overhead, Aguayo, Bicket *et al.* [15] assumed 1500 Bytes as a constant size of the packet and suggested Estimated Transmission Time metric is defined as:

$$EstdTT(l) = \frac{1}{reliability(l) * rate(l)} \quad (18a)$$

This can alternatively be written using (2), as:

$$EstdTT(l) = \frac{ETX(l)}{rate(l)} \quad (18b)$$

### 13. ETX Distance

*ETX* metric is combined with greedy forwarding to optimize routing path without relying the frequently broadcast route probing messages (as in original $ETX$) in [16]. $ETX$ virtual distance between pairwise nodes $x_i$ and $x_j$ as the minimal $ETX$ among all the routing paths connecting $x_i$ and $x_j$, i.e.,

$$\delta(x_i, x_j) = \min_{l_i \in L} ETX(l_i) \quad (19)$$

| Metric | Performance(s) over ETX | Target Platform(s) | Design Requirement(s) |
|---|---|---|---|
| ETX | | Multi-Hop Wireless Networks | Maximizing throughput |
| Modified ETX | 1.Accuracy of loss estimation. 2.50% reduction in average packet loss. | Time-Varying WMN's | Selecting good paths in WMN's among the time-varying binary symmetric channel |
| ENT | 1.Accuracy of loss estimation. 2.50% reduction in average packet loss least link layer transmissions. | Time-Varying WMN's | Selecting good paths in WMN's among the time-varying binary symmetric channel |
| ETT | Only summming-up the links ignores intra-flow interference (on the same channel) | Multi-Radio, Multi-Hop WMN's | Measuring the loss rate and the bandwidth |
| WCETT | Multiple Channels | Multi-Radio, Multi-Hop WMN's | Choosing a high-throughput path between a source and a destination |
| MIC | It Captures both inter-flow and intra- flow interference. | Multi-hop Wireless Networks | Enabling efficient calculation for minimum weight path and loop-free routing |
| iAWARE | It considers intra-flow and inter-flow interference, medium instability, and data-transmission time. | Multi-Radio WMN's | Finding paths that are better in terms of reduced interflow and intra-flow interference for multi-radio networks. |
| DBETX | 1.Improves network performance in the presence of varying channels. 2.Outperforms ETX in the presence of fading. | Fading Wireless Channels | Maximize the through on the Fading Channels |
| EETT | Select multi-channel routes with the least interference to maximize the end-to-end throughput. | Large-Scale Multi-Radio Mesh Networks | Maximizing end-to-end throughput |
| EDR | 1.Finds transmission contention degree of each link as a function of the wireless link loss. 2.Quantifies the impact of the wireless link loss on medium access backoff. 3.Considers possible concurrent transmissions when two links do not interfere with each other. | Multi-Hop Wireless Networks | Finding high-throughput paths in multi-hop ad hoc wireless networks. |
| ETP | More Accurate Throughput Estimations. | Multi-rate Multi-channel Mesh Networks | Maximizing throughput |
| MCR | It simplifies the use of multiple channels by using multiple interfaces | Multi-Channel Multi-Interface Ad hoc Wireless Networks | To increase the available network capacity using multiple channels |
| MTM | 1.MAC-related overhead. 2.Gives higher throughput, 3. Selects more reliable links | Multi-rate Ad hoc Wireless Networks | avoiding the long range links often selected by shortest path |
| EstdTT (SrcRR Algorithm) | 1.Predicts best 802.11 transmission bit-rate on each link. 2.Reduces loss rate avoiding TCP's timeouts and idle time and improves the choice of link transmission bit-rate. | 802.11 Mesh Networks | Improving varying link loss rates, transient bursts of losses, poor transmit bit-rate selection, failure to identify high throughput routes, |
| ETX Distance | Greedy forwarding based on ETX Distance outperforms previous geographic routing approaches. | Geographical WSN's | Greedy forwarding |
| Multicast ETX | Improves throughput or reliability, possibly at the cost of minimum energy consumption. | Multi-hop Wireless Networks | Energy-efficient reliable communication in the presence of unreliable or lossy wireless link |

Table 1. Performances over ETX, design goals and platforms of the ETX based metrics

Where $L$ is the set of hops or paths connecting $x_i$ and $x_j$. It has been suggested that $ETX$ distances between pairwise nodes in a WSN can be inferred from their virtual coordinates. Making the comparison of the $ETX$ distances between neighboring nodes, the greedy forwarding can determine the next hop. $ETX$ distance comparison based greedy forwarding guides a packet towards the correct direction and deliver the packet through consecutive hop by hop forwarding, as $ETX$ distance directly reflects the length of a communication path between pairwise nodes in a WSN.

## 14. Multicast ETX (METX)

This energy-efficient routing metric [17], aims to minimize the total transmission energy, in the presence of an unreliable link layer, for the path:

$$C(s,d) = \frac{C(s,u) + W(u,d)}{1 - p_{el}} \quad (20)$$

$C(s,d)$ is the expected energy-cost of transmission from a source $s$ to destination $d$, $l$ is the link between $u$ and $d$ in the path, $p_{el}$ is the error rate of that link, and $W(u,d)$ is the transmission energy required between nodes $u$ and $d$. The authors in [18] modified the metric given by Eq. (20) to a new metric, $METX$ setting $W(u,d)$ to 1, as WMN's are not energy sensitive. Eq. (21a) gives us the total expected number of transmissions needed by all the nodes along a path from a source to a destination in order to guarantee successful reception of at least one packet at the receiver:

$$METX = \sum_{i=1}^{n} \frac{1}{\prod_{j=i}^{n}(1-p_{ej})} \quad (21a)$$

In terms of *ETX* using Eq (1):

$$METX = \sum_{i=1}^{n} \frac{1}{\prod_{j=i}^{n} \frac{1}{ETX_j}} \quad (21b)$$

*i* denotes the *ith* link from a source to a destination comprising *n* links.

## III. CONCLUSION

After minimum hop count which usually selects lossy links, *ETX* is the most widely used routing link metric (in the presence of least mobility of nodes and availability of links). We therefore, analyzed and compared the performance of those wireless routing which are *ETX* based and are used by the recent routing protocols. Overheads occurred and throughputs achieved due to the factors added to *ETX* have been listed and discussed. Future work goals are to simulate these metrics with the most widely used protocols, as DSR, AODV, OLSR, etc and to analyze their performance over *ETX* and minimum hop count.